\documentstyle[12pt,epsfig]{article}
\begin{document}
\title{Non--Newtonian gravity and coherence properties of light.}
\author{ A. Camacho
\thanks{email: acamacho@nuclear.inin.mx} \\
Department of Physics, \\
Instituto Nacional de Investigaciones Nucleares\\
Apartado Postal 18--1027, M\'exico, D. F., M\'exico.}

\date{}
\maketitle

\begin{abstract} 
In this work the possibility of detecting a non--Newtonian contribution to the gravitational potential by means of its effects upon the first and second--order coherence properties of light is analyzed. It will be proved that, in principle, the effects of a fifth force upon the correlation functions of electromagnetic radiation could be used to detect the existence of new forces. Some constraints upon the experimental parameters will also be deduced. 
\end{abstract}

\section{Introduction}

Optical interferometry has played a fundamental role in some experimental aspects of gravitational physics [1], for instance, we may mention that there are gravity--waves detectors which are built following Michelson interferometer [2], or that Sagnac ring interferometer [3] constitutes the bedrock for the so--called ring laser gyroscopic device, the one could be used to test the different metric theories of gravity in the weak--field and slow motion limit [4].

It is also needless to say that general relativity (GR) is one of the milestones of modern physics, and that nowadays many of its predictions have been already confronted against some experiments [5]. Nevertheless, we may find several theore\-tical attempts to construct a theory of elementary particles, which naturally predict the existence of new forces (usually refered as fifth force) whose effects extend over macroscopic distances [6].
A crucial characteristic of these forces is that they are not described by an inverse--square law, and even more, they, generally, violate the Weak Equivalence Principle (WEP) [7]. 

After more than a decade of experiments [8], there is no compelling evidence for any kind of deviations from the predictions of Newtonian gravity. Neverwithstanding, Gibbons and Whiting (GW) phenomenological analysis of gravity data [9] has proved that the very precise agreement between the Newtonian gravity and the observation of planetary motion does not preclude the existence of large non--Newtonian effects over smaller distance scales, i.e., precise experiments over one scale do not necessarily constrain gravity over another scale. GW conclusions allowed them to affirm that the current experimental constraints over possible deviations did not severly test Newtonian gravity over the $10$--$1000$m distance scale, usually denoted as the ``geophysical window".

Looking at the experimental efforts that have been done in order to test the inverse--square law we will find that they can be separated into two large classes: (i) those experiments which involve the direct measurement of the magnitude of $G$ [10]; and (ii) the direct measurement of the magnitude of $G(r)$ with $r$ [11]. At this point it is noteworthy to comment that recently some new proposals have been considered [12], which do not fall in these aforementioned two cases.

In the present work we will analyze the possibility of detecting a Yukawa--type contribution to the gravitational potential (which is one of the possibilitites in this direction [13]) considering the effects of this fifth force upon the first and second--order coherence properties of light. In other words, we will consider a Young--type experi\-ment with light from two atoms [14], and evaluate the effects of a non--Newtonian gravitational term on the resulting first--order correlation function. Also the consequences upon the Hanbury--Brown--Twiss effect [15] of these kind of terms will be consi\-dered. Some possible experimental scenarios will also be shown.
\bigskip
\bigskip

\section{Young's experiment and non--Newtonian gra\-vity}
\bigskip
\bigskip

Let us consider two identical atoms (located at $P$ and $P'$), where each one of them has two levels, and a single photon, such that only one of these atoms will be excited.
The initial state of our system reads

\begin{equation}
a(\left\vert 0,1'> + \vert 1,0'>\right)\vert \tilde 0> + b\vert 0, 0'>\vert \phi>.
\end{equation}

Here $\vert 0>, \vert1>, \vert 0'>, \vert 1'>$ denote the ground and excited states of the two atoms, while $\vert \tilde 0>$ is the vacuum of the electromagnetic field, and $\vert\phi>$ designates the photon. After a time larger than the mean decay time,$t_m$, the system decays to 

\begin{equation}
\vert \alpha> = {1\over\sqrt{2}}\vert 0,0'>\left[\vert\gamma> + \vert\gamma'>\right].
\end{equation}

In this last expression $\vert\gamma>$ and $\vert\gamma'>$ denote the photon states emited from sites $P$ and $P'$, respectively.

Let us now assume that the the gravitational interaction contains a Yukawa--type term [13]

\begin{equation}
V(r) = -{G_{\infty}mM\over r}\left[1 + \alpha\exp{(-r/\lambda)}\right].
\end{equation}

$G_{\infty}$ describes the interaction between $M$ and $m$ in the limit case $r\rightarrow\infty$, i.e., $G_N = G_{\infty}(1 + \alpha)$, where $G_N$ is the Newtonian constant [8]. We may understand this kind of deviation term as a consequence of the exchange of a single new quantum of mass $\tilde m$, $\lambda = {\hbar\over c\tilde m}$, this field is usually denoted dilaton.

Hence the gravitational potential generated by $M$ reads

\begin{equation}
U(r) = -{G_{\infty}M\over r}\left[1 + \alpha\exp{(-r/\lambda)}\right].
\end{equation}

The interference experiment will detect at point $S$ the light that results from the decay of the system. But here we will take into account the redshift in the frequency that appears as a consequence of the fact that the electromagnetic field {\it climbs} in a region where a non--vanishing gravitational field is present. In other words, if the frequency at the emission point is $\nu$, and the radiation is detected at a point, which respect to the emission point has a difference $\Delta U$ in the gravitational potential, then the frequency at the detection point reads [5]

\begin{equation}
\tilde{\nu} = {\nu\over 1 + \Delta U/c^2}.
\end{equation}

As is already known the electromagnetic field operator can be separated into two parts, namely, with positive and negative frequency parts [16]. Nevertheless, in the case of an experiment which employs absorptive detectors the measurements are destructive, and in consequence only that part of the field operator containing annihilation operators, ${\bf E}^{(+)}({\bf r}, t)$, has to be considered. In order to simplify the model we will assume that the field is linearly polarized, and that the radiation emitted from $P$ (or $P'$) is monocromatic.

One of the ideas behind this proposal is to consider the possibility of performing this kind of experiment near the Earth's surface, hence we will assume that

\begin{equation}
r = R + z,
\end{equation}

where $R >>\vert z\vert$.

Under these conditions the field operator containing the annihilation operator reads, approximately

\begin{equation}
E^{(+)}({\bf r}, t) = \Xi\hat{a}\exp\left\{-i\nu\left[1 - {g_0\over c^2}h
{1 + \alpha e^{(-R/\lambda)}\over 1 + \alpha}\right]\left[t - \hat{k}\cdot{\bf r}\right]\right\}.
\end{equation}

Here $h$ is the {\it climbed} distance, $\hat{k}$ denotes the unitary vector in the direction of propagation, $\Xi$ is a constant with dimensions of electric field, $\hat{a}$ is the corresponding annihilation operator, and $g_0 = g_{\infty}(1 + \alpha)$ is the effective acceleration of gravity at laboratory distances.

The first--order correlation function is given by [14] 

\begin{equation}
G^{(1)}({\bf r}, {\bf r}; t, t) = <\alpha\vert E^{(-)}({\bf r}, t)E^{(+)}({\bf r}, t)\vert \alpha>.
\end{equation}

The radiation stemming from $P$ (and also from $P'$) has to be described, according to the rules of quantum theory, as a superposition of plane wave states [14], ne\-verwithstanding, we may suppose, without introducing unphysical assumptions, that the state vector of $\vert\gamma>$ (and, of course, of $\vert\gamma'>$) is given by a plane wave [16]. 
In this particular case, and remembering that $h$ and $h'$ are the {\it climbed} distances coming from $P$ and $P'$, respectively, we have that 

\begin{equation}
G^{(1)}({\bf r}, {\bf r}; t, t) = \vert \Xi\vert^2\left\{1 + \cos\left([{\bf k} - {\bf k'}]\cdot{\bf r} + \tilde{g}[h{\bf k} - h'{\bf k'}]\cdot{\bf r} + \tilde{g}\nu t\Delta h\right)\right\}.
\end{equation}

At this last expression we have introduced the following definition

\begin{equation}
\tilde{g} = {g_0\over c^2}{1 + \alpha e^{(-R/\lambda)}\over 1 + \alpha}.
\end{equation}

Where we have that $\Delta h = h' - h$.
It is also readily seen that if $g_0 = 0$, then we recover the usual Young's interference pattern [16].
\bigskip
\bigskip

\section{Interference patterns}
\bigskip
\bigskip

At this point we must mention that a remarkable difference of expression (9) with respect to the case in which gravity is absent concerns the explicit time dependence of the interference pattern, a fact that can be understood noting that in the case without gravitational field time 
disappears from the corresponding expression because both waves do have the same frequency [16], a fact that in our case does not happen, indeed the difference in the {\it climbed} distance renders different frequencies.
\bigskip
\bigskip

\subsection{Time independent interference pattern}
\bigskip
\bigskip

Concerning expression (9) a possibility comprises the case in which $h = h'$, i.e., $\Delta h = 0$. In this case expression (9) may be rewritten as

\begin{equation}
G^{(1)}({\bf r}, {\bf r}; t, t) = \vert \Xi\vert^2\left\{1 + \cos\left(A\left[1 + \tilde{g}(h + h')\right]\right)\right\}.
\end{equation}

$A$ is a factor present in the case in which there is no gravitational field, and it depends upon the geometry of the inteferometer, and also on the wavelength of the emitted radiation [16].

If we try to detect the effects of a fifth force inside the so--called ``geophysical window" [9], we may consider the following values $\alpha \in [10^{-3}, 10^{-2}]$ and  $\lambda = 10$m [17]. From expression (10), and remembering that $g_0 = G_0M/R^2$ we obtain a condition on $R$ as function of $h$ and $h'$. For instance, if $\tilde{g}[h + h']\sim 10^{-8}$, then 

\begin{equation}
(h + h')/R^2\sim 10^{-4}m.
\end{equation}
\bigskip
\bigskip

\subsection{Time dependent interference pattern}
\bigskip
\bigskip

From expression (9) we may see that there are certain time values, $t_n$ ($n\in N$), such that
\begin{equation}
\tilde{g}\Delta h\nu t_n = 2\pi n.
\end{equation}

 Hence the interval between $t_{n+1}$ and $t_n$ is

\begin{equation}
\Delta t_n \equiv t_{n+1} - t_{n} = {2\pi\over \tilde{g}\nu\Delta h}.
\end{equation}

If we consider this last expression for the purely Newtonian case

\begin{equation}
\Delta t_n^{(N)} = {2\pi c^2\over g_0\nu\Delta h},
\end{equation}

and compare it against the non--Newtonian situation

\begin{equation}
\Delta t_n^{(NN)} = {2\pi c^2\over g_0\nu\Delta h}{1 + \alpha\over 1 + \alpha e^{-R/\lambda}},
\end{equation}

we deduce that 

\begin{equation}
\Delta t_n^{(NN)}/\Delta t_n^{(N)} = {1 + \alpha\over 1 + \alpha e^{-R/\lambda}}.
\end{equation}

Employing the aforementioned values for $\alpha$ and $\lambda$ we have, approximately, that

\begin{equation}
\Delta t_n^{(NN)}/\Delta t_n^{(N)} = 1 + 10^{-3},
\end{equation}

In this context the possibility of detecting a {\it fifth force} depends on the condition that the difference between the non--Newtonian and Newtonian cases has to be larger than the resolution of the experimental apparatus, i.e., $\vert\Delta t_n^{(NN)}- \Delta t_n^{(N)}\vert > \Delta T$, where $\Delta T$ denotes the time resolution of the measuring device. 

In order to have a realistic experimental situation we must take into account the fact that the emitted radiation is really a pulse. The lifetime of this pulse, assuming that the radiative energy loss (of the dipole oscillator that acts as emitter) is very slow compared with a period of atomic dipole oscillation, has an order of magnitude of $\tau \sim 0.1\mu$s [18].
The possibility of detecting time intervals of $50$fs, based on the interference of two--photon probability amplitudes in two--photon detection [19], implies that the aforementioned differences ($\sim 0.1\mu$s) could represent no technological difficulty.

As was mentioned before, one possibility is to consider the radiation within the optical spectrum, hence we may introduce the following wavelength, for the emitted field, $\lambda^{(r)}\sim 400$nm. Therefore, if $\Delta t_n^{(NN)}\sim 0.01\mu$s, then we obtain, from (16), a constraint upon $\Delta h$, as function of $R$

\begin{equation}
\Delta h/R^2 \sim 10^{-4}m^{-1}.
\end{equation}

If the experiment were performed near the Earth's surface ($R\sim 10^6$m), then $\Delta h \sim 10^4$m.
\bigskip
\bigskip

\section{Hanbury--Brown--Twiss effect}
\bigskip
\bigskip

Let us now consider the consequences of a fifth force of Yukawa--Type upon the so called Hanbury--Brown--Twiss effect (HBT) [15], i.e., we must now analyze the second--order coherence properties of light.

Once again we have two atoms, located at points $P$ and $P'$, but now there are two detection points, $S_1$ and $S_2$. Initially the atoms are excited, but there is no electromagnetic field, hence the initial state vector reads

\begin{equation}
\vert\alpha(t = 0)> = \vert 1,1'>\vert\tilde{0}>.
\end{equation}
 
After an interval much larger than the atomic decay time, $t_m$, the system becomes

\begin{equation}
\vert\alpha(t >>t_m)> = \vert 0,0'>\vert\gamma, \gamma'>.
\end{equation}

Resorting to the definition of second--order correlation function [15], we find, a\-ssuming once again the plane wave approximation for the emitted radiation, that the interference term is given by

\begin{equation}
\cos\left\{\left[{\bf k} - {\bf k'}\right]\cdot\left[{\bf r_2} - {\bf r_1}\right] + \tilde{g}[h'_2{\bf k'} - h_2{\bf k}]\cdot{\bf r_2} - \tilde{g}[h'_1{\bf k'} - h_1{\bf k}]\cdot{\bf r_1} + \tilde{g}\nu t\left[\Delta h - \Delta h'\right]\right \}.
\end{equation}

Here we have that $h_1$ and $h_2$ are the {\it climbed} distances for the radiation emitted by $P$ and detected at $S_1$ and $S_2$, respectively (we have the same argument if the light is emitted in $P'$). Additionally, $\Delta h= h_2 - h_1$, $\Delta h'= h'_2 - h'_1$. Imposing the condition $g_0 = 0$, we recover the usual HBT situation [15].
\bigskip
\bigskip

\subsection{Experimental possibilities}
\bigskip
\bigskip

Comparing (22) with the argument of the cosine function in (9) we may see that the order of magnitude of some of the experimental parameters, for instance, $R$ or $t$, that emerge in connection with HBT in the context of the so called ``geophysical window", will be the same as in the analysis of a Young's experiment. 

The additional time independent terms in (22) have the same structure as the corresponding one in (9), which means that if in the HBT case we impose the condition $\tilde{g}\vert[h'_2{\bf k'} - h_2{\bf k}]\cdot{\bf r_2}\vert\sim 10^{-8}$, then 

\begin{equation}
\vert\Delta h'\vert/R^2\sim 10^{-4}m^{-1},
\end{equation}

here appears $\vert\Delta h'\vert$, and not $h + h'$, as in (12).

Additionally, instead of (14), we have that 

\begin{equation}
\Delta t_n = {2\pi\over \tilde{g}\nu\vert\Delta h - \Delta h'\vert}.
\end{equation}

In other words, if we go from Young's experiment to HBT, then $\Delta h$ is replaced by $\vert\Delta h - \Delta h'\vert$.

\section{Conclusions}
\bigskip
\bigskip

We have considered the effects, on the first and second--order coherence properties of light, of a Yukawa--type modification to the gra\-vitational potential. We have seen, expressions (9) and (22), that the resulting interference patterns do depend on time, i.e., we do not deal with statistically stationary fields. 
This last feature is a consequence of the fact that the appearing radiation beams have different redshifts, they {\it climb} different distances. 
Resorting to this time dependence, the conditions on some of the experimental parameters, that could lead to the detection of this Yukawa--type modification, have also been addressed. Roughly stated, if the experiment is carried out in the optical region, then the distance {\it climbed} by the radiation has to be two orders of magnitude smaller than the distance existing from the experimental device to the Earth's center, see expressions (12) and (19). 

This last statement also allows us to reduce the order of magnitude of the {\it climbed} distance, namely, we may change the value of $R$, i.e., modify the distance from the Earth's center to the point where the experimental device is located (for example, carrying out the experiment inside the Earth, in a mine), and in consequence we may see (expressions (12) and (19)) that the order of magnitude of $\Delta h$ (or $h$ and $h'$) would also decrease. 

Clearly, the present idea has to be improved in order to have a feasible expe\-rimental proposal. One possibility in this direction comprises the possible increase of the effective value of $\Delta h$ (or of $\vert\Delta h - \Delta h'\vert$) by the use of some kind of {\it compact interferometric detector}, and idea already considered in the detection of gravitational waves [20], the one allows very long optical paths without needing interferometric devices with very large arm lengths. \bigskip
\bigskip

\Large{\bf Acknowledgments}\normalsize
\bigskip

The author would like to thank A. A. Cuevas--Sosa for his 
help and J.Cervantes--Cota for the fruitful discussions on the subject. This work was partially supported by CONACYT (M\'exico) Grant No. I35612--E.

\end{document}